\documentclass[prl, superscriptaddress, showpacs, twocolumn]{revtex4-1}
\usepackage{amsmath}
\usepackage{amssymb}
\usepackage{graphicx}
\usepackage{bm}
\usepackage{color}

\begin{document}

\title{From the Cooper problem to canted supersolids in Bose-Fermi mixtures}
\author{Peter Anders}
\affiliation{Theoretische Physik, ETH Zurich, 8093 Zurich, Switzerland}
\author{Philipp Werner}
\affiliation{Department of Physics, University of Fribourg, 1700 Fribourg, Switzerland}
\affiliation{Theoretische Physik, ETH Zurich, 8093 Zurich, Switzerland}
\author{Matthias Troyer}
\affiliation{Theoretische Physik, ETH Zurich, 8093 Zurich, Switzerland}
\author{Manfred Sigrist}
\affiliation{Theoretische Physik, ETH Zurich, 8093 Zurich, Switzerland}
\author{Lode Pollet}
\affiliation{Theoretische Physik, ETH Zurich, 8093 Zurich, Switzerland}
\affiliation{Department of Physics and Arnold Sommerfeld Center for Theoretical Physics, Ludwig-Maximilians-Universit{\"a}t M{\"u}nchen, D-80333 M{\"u}nchen, Germany}

\date{\today}

\begin{abstract}

We calculate the phase diagram of the Bose-Fermi Hubbard model on the $3d$ cubic lattice at fermionic half filling and bosonic unit filling by means of single-site dynamical mean-field theory (DMFT). For fast bosons, this is equivalent to the Cooper problem in which the bosons can induce s-wave pairing between the fermions.  We also find miscible superfluid and canted supersolid phases depending on the interspecies coupling strength. 
%The modifications in the phase diagram compared to the purely fermionic case are well explained by renormalized hoppings and on-site interactions. 
In contrast, slow bosons favor  fermionic charge density wave structures for attractive fermionic interactions. These competing instabilities lead to a rich phase diagram within reach of cold gas experiments.

\end{abstract}

\pacs{71.10.Fd}

\hyphenation{}

\maketitle

%\section{Introduction}
Interactions between bosons and fermions play a crucial role in various physics contexts. Examples include the atomic nucleus, quarks exchanging gluons via the strong force, electrons dressed by lattice vibrations forming polarons, conventional superconductors where phonons induce an attraction between the electrons at the Fermi energy, and the phase separation between $^3$He and $^4$He mixtures. Beyond mean-field, these systems are notoriosly difficult to describe. Cold atom experiments can be used to simulate this physics, thanks to the experimental control over the coupling strength between fermions and bosons, effectively performing quantum simulation of superconductors. 

The first experiments investigated the influence of fermions on the bosonic Mott insulator, and found that the bosonic visibility always decreases when adding fermions attractively interacting with the bosons~\cite{Ospelkaus06, Guenter06}. This has been explained by self-trapping~\cite{Ospelkaus06, Luehmann08, Best08, Tewari09}, corrections to higher bands~\cite{Tewari09, Mering08}, or by adiabatic heating~\cite{Guenter06, Pollet08,  Cramer08}. At weaker inter-species interactions, symmetry between repulsion and attraction was found~\cite{Best08}. In a dynamics experiment the strength of the potential terms has been measured with astonishing precision~\cite{Will11}. However, many more exotic phases such as supersolids~\cite{Buechler03} and pair superfluids~\cite{Kuklov03} have been predicted~\cite{Albus03, Illuminati04, Lewenstein04, Mathey04, Cramer04, Pollet06, Titvinidze08, Titvinidze09}, though not yet realized in experiment. Such may become possible though thanks to the recent discovery of multiple Feshbach resonances between $^{23}$Na and  $^{40}$K at MIT ~\cite{Zwierlein12}.

In this Letter, we revisit the Cooper problem of conventional superconductors in a cold atom setup, that is we study the conditions under which bosons induce s-wave pairing between spin-1/2 fermions~\cite{Bardeen67, Bijlsma00, Viverit}. We will see that a bosonic condensate leads to a strong static enhancement of s-wave pairing.
Our formalism also allows us to explore physics in the strong Bose-Fermi coupling regime as well as bosons that are slow compared to the Fermi velocity. In such cases, instabilities favoring density waves compete against pairing, leading to a rich and unexpected phase diagram.
 % via the density-density coupling between bosons and fermions. % and the $SU(2)\times SU(2)$ symmetry of the pure fermions at half filling. \\

Our model consists of spinless bosons and spin-$1/2$ fermions on a cubic lattice with Hamiltonian
\begin{align}
&H = - t_{\rm f} \sum_{\langle ij \rangle \sigma}  c_{i\sigma}^{\dagger} c_{j\sigma} -  t_{\rm b} \sum_{\langle ij \rangle} b_i^{\dagger} b_j  - \mu_{\rm f}  \sum_{i\sigma} n_{i\sigma}^{\rm f} - \mu_{\rm b} \sum_i n_i^{\rm b} \nonumber \\
&\hspace{3mm}+U_{\rm ff}\sum_i n_{i\uparrow}^{\rm f} n_{i\downarrow}^{\rm f} + \frac{U_{\rm bb}}{2} \sum_i n_i^{\rm b} (n_i^{\rm b}-1) + U_{\rm bf} \sum_{i\sigma} n_i^{\rm b} n_{i\sigma}^{\rm f}  \nonumber,
\end{align}
where $b^{\dagger}_i$ and $b_i$ ($c^{\dagger}_{i\sigma}$ and $c_{i\sigma}$) are the bosonic (fermionic) creation and annihilation operators at site $i$ with spin $\sigma$ and $n_i^{\rm b}$ ($n_i^{\rm f}$) denote the corresponding number operator. Particles can hop between neighboring sites via the hopping amplitude $t_{\rm b(f)}$ and the particle number is adjusted through the chemical potential $\mu_{\rm b(f)}$. The particles can interact via an onsite interaction, where $U_{\rm bb}$, $U_{\rm ff}$ and $U_{\rm bf}$ denotes the boson-boson, fermion-fermion and boson-fermion interaction, respectively. We will work at unit filling for the bosons and half filling for the fermions, in which case the sign of $U_{\rm bf}$ is irrelevant. This model is a direct extension of the previous cold atom experiments with spin-polarized fermions. We restrict the discussion to the case where the spin-up and spin-down fermions interact equally strongly with the bosons.
% (the interspecies coupling may be spin-dependent when using a Feshbach resonance).

To numerically study the above model we use DMFT where the full many body problem is mapped onto a self-consistent determination of an impurity model. In the Nambu notation the kinetic impurity action for sublattice $s$ is given by

\begin{align}
%&S=- \frac{1}{2}\int_0^{\beta}\hspace{-1.5mm} d\tau d\tau' \textbf{b}^{\dagger}(\tau) \mathbf{\Delta}_{b,s}(\tau-\tau') \textbf{b}(\tau') - \kappa \mathbf{\Phi}^{\dagger} \int_0^{\beta}\hspace{-1.5mm} d\tau  \textbf{b}(\tau) \nonumber \\
   & S^{\rm kin}_s=- \frac{1}{2}\int_0^{\beta}\hspace{-1.5mm} d\tau d\tau' (\textbf{b}_s^{\dagger} (\tau) - \mathbf{\Phi}_s^{\dagger} ) \mathbf{\Delta}_{{\rm b},s} (\tau-\tau') ( \textbf{b}_s(\tau') - \mathbf{\Phi}_s )   \nonumber \\
& - zt \mathbf{\Phi}_{-s}^{\dagger} \int_0^{\beta} d\tau  \textbf{b}_s(\tau) - \int_0^{\beta}\hspace{-1.5mm} d\tau d\tau' \textbf{c}_s^{\dagger}(\tau) \mathbf{\Delta}_{{\rm f},s} (\tau-\tau') \textbf{c}_s(\tau')\nonumber %\\
% &\hspace{6.5mm}+ \int_0^{\beta}\hspace{-1.5mm} d\tau H_{\text{loc}}(\tau),
%\label{action}
\end{align}
where $\mathbf{\Delta}_{\rm b (f)}$ is the matrix hybridization function of the bosons (fermions) and the corresponding creation and destruction operators are given by $\mathbf{b}_s^{\dagger}(\tau)=(b^{\dagger}_s(\tau), b_s(\tau))$ and $\mathbf{c}^{\dagger}_s(\tau)=(c_{\uparrow, s}^{\dagger}(\tau),c_{\downarrow, s}(\tau))$, and $s=A,B$ denote the two sublattices.
$\mathbf{\Phi}_{-s}^{\dagger}= \langle \mathbf{b}^{\dagger} \rangle_{-s} = (\phi_{-s}^*, \phi_{-s})$ is the time independent condensate order parameter of the bosons determined selfconsistently on the other sublattice as denoted by the subscript $-s$. 
For a cubic lattice, the coordination number is $z=2d=6$. 
The fermionic hybridization function is determined by the following form of the inverse lattice Green function [$\mathbf{c}^{\dagger}(\tau)=(c_{\uparrow, A}^{\dagger}(\tau),c_{\downarrow, A}(\tau),  c_{\uparrow, B}^{\dagger}(\tau), c_{\downarrow, B}(\tau))$]
\begin{align}
\bm{G}_{\rm f}^{-1}( \bm{k}, i\omega_n) =
\begin{bmatrix}
\zeta - \Sigma_{A} & -\tilde{\Sigma}_A & - \epsilon_k & 0 \\
-\tilde{\Sigma}_A & -\zeta^* + \Sigma_A^* & 0 & \epsilon_k \\
-\epsilon_k & 0 & \zeta - \Sigma_B & -\tilde{\Sigma}_B \\
0 &  \epsilon_k & - \tilde{\Sigma}_B & -\zeta^* + \Sigma_B^*
\end{bmatrix}
\label{eq:Nambu_spinor}
\nonumber
\end{align}
(with  $\zeta = i\omega_n + \mu, \epsilon_k = 2t_{\rm f} \sum_{j=1}^d \cos (k_j)$, and standard notation for the normal and anomalous selfenergies on the respective sublattices) such that (charge) density wave ordering and s-wave pairing are allowed, and can occur independently of each other. The nature of the density-density coupling between bosons and fermions implies that a density wave ordering for fermions immediately creates density wave ordering for the bosons, and vice versa.  The possible symmetry breakings in the spin sector are expected to be the same as in the pure fermionic model.
%The local part of the action is given by
%
%\begin{align}
%&\hspace{-2mm}H_{\text{loc}}(\tau) = -\mu_b n^b(\tau) -\mu_f\sum_{\sigma} n_{\sigma}^f(\tau) + U_{ff} n_{\uparrow}(\tau)n_{\downarrow}(\tau) \nonumber \\
%&\hspace{9.5mm}+ \frac{U_{bb}}{2} n^b(\tau)[n^b(\tau)-1] + U_{bf}\sum_{\sigma} n^b(\tau)n_{\sigma}^f(\tau).
%\end{align}
%
The (local) potential energy terms are absorbed in the potential part of the impurity action $S_{\rm pot} = \int_0^{\beta}\hspace{-1.5mm} d\tau H_{\text{loc}}(\tau)$. %$H_{\rm loc}$. 
%We write the matrix of the hybridization function in the Nambu notation as
%%
%\begin{equation}
%\mathbf{\Delta}_{\rm b} (\tau) \hspace{-0.5mm}=\hspace{-0.5mm} \left( \hspace{-1mm}\begin{array}{cc}
%F_{\rm b} (-\tau)  & \hspace{-1mm}2K_{\rm b} (\tau) \\
%2K_{\rm b}^*(\tau) & \hspace{-1mm}F_{\rm b} (\tau)
%\end{array} \hspace{-1mm}\right)\hspace{-1mm}, \mathbf{\Delta}_{\rm f} (\tau)\hspace{-0.5mm} =\hspace{-0.5mm}  \left(\hspace{-1mm} \begin{array}{cc}
%-F_{\uparrow}(-\tau)  & \hspace{-1mm}K_{\rm f}(\tau) \\
%K_{\rm f}^*(\tau) & \hspace{-1mm}F_{\downarrow}(\tau)
%\end{array}\hspace{-1mm} \right)\hspace{-1mm}. \nonumber
%\end{equation}
%%
%and the elements of the Green's function as
%%
%\begin{equation}
%\mathbf{G}_f(\tau) = -\langle T \mathbf{c}(\tau)\mathbf{c}^{\dagger}(0)\rangle = \left( \begin{array}{cc}
%G_{\uparrow}(\tau)  & \tilde{G}(\tau) \\
%\tilde{G}^*(\tau) & G_{\downarrow}(\tau)
%\end{array} \right),
%\end{equation}
%where $T$ is the time ordering operator, $G_{\uparrow}(\tau) = -\langle T c_{\uparrow}(\tau)c_{\uparrow}(0)\rangle$ denotes the normal Green's function and $\tilde{G}(\tau) = -\langle T c_{\uparrow}(\tau)c_{\downarrow}(0)\rangle$ and $\tilde{G}^*(\tau) = -\langle T c^{\dagger}_{\downarrow}(\tau)c^{\dagger}_{\uparrow}(0)\rangle$ anomalous Green's functions. From the anomalous Green's function we directly obtain the superfluid order parameter $\Delta^{\rm BCS} = \langle c_{\uparrow} c_{\downarrow} \rangle= -\tilde G(0_{+}).$
%
%\begin{equation}
%\Delta = \langle c_{\uparrow} c_{\downarrow} \rangle= -\tilde G(0_{+}).
%\end{equation}

\begin{figure}
\begin{center}
\includegraphics[angle=0, width=0.8\columnwidth]{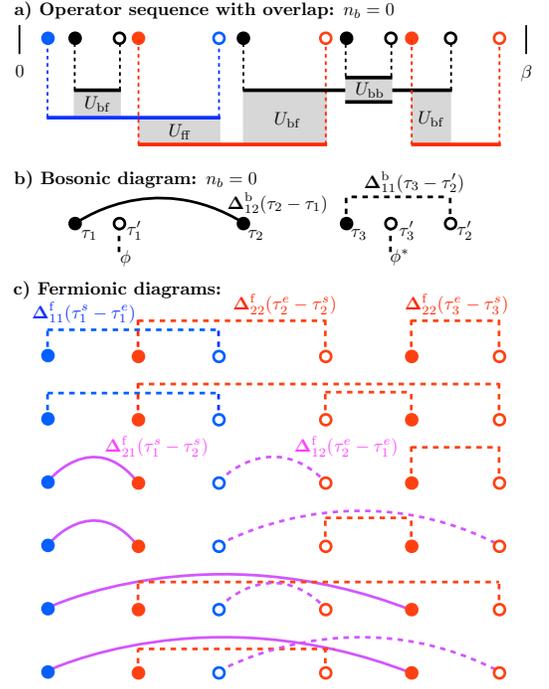}
\caption{(Color online) Illustration of a typical Monte Carlo configuration. The full (empty) circles denote creation (annihilation) operators in the imaginary time interval $[0, \beta)$ for bosons (black), and spin-up (blue) and spin-down (red) fermions. a) The local contribution to the weight of the operator sequence is determined by the length of the segments and the overlap between segments of different particles (segments mark time intervals in which a particle resides on the impurity). b) A possible configuration of bosonic hybridization functions and source fields with density $n_{\rm b} = 0$ at imaginary time $\tau = 0$. c) All possible combinations to connect the fermionic creation and annihilation operators. 
%The weight of these terms can be summed up into the determinant of a (fermionic) hybridization matrix. 
%In the simulations random configurations with all possible pairings of the types illustrated here are generated according to their respective weights.
}
\label{Fig_diagrams}
\end{center}
\end{figure}

As impurity solver we use a continuous-time Monte Carlo method based on an expansion of the partition function in powers of the impurity-bath hybridization $\mathbf{\Delta}_{\rm b(f)}$ and the condensate order parameter $\mathbf{\Phi}$. The method is a direct extension of the fermionic~\cite{Werner06} and bosonic~\cite{Anders10, Anders11} impurity solvers.
%(extended to anomalous hybridization functions, to treat the superfluid state of the fermions). 
This method allows for the first time to study Bose-Fermi mixtures within the full DMFT formalism (~\cite{Byczuk11}, see however Refs.\cite{Anders10, Anders11} regarding the broken symmetry in the action). An illustration of a possible Monte Carlo configuration is shown in Fig.~\ref{Fig_diagrams}. Details of the algorithm will be presented elsewhere~\cite{Anders_PhD}.

%\section{Results}
%In this section we present our results for the Bose-Fermi Hubbard model . 
%All simulations were performed at half filling for the fermions $(n_{\uparrow}=n_{\downarrow}=1/2)$ and unit filling $(n_b=1)$ for the bosons on the $3d$ cubic lattice. First, we will treat the bosons in the static mean-field approximation $(\mathbf{\Delta}_b(\tau)=0)$ in order to avoid the sign problem. We will then show that the difference between this approximation and the full B-DMFT is negligible if the condensate fraction of the bosons is large. However, close to the phase transition of the bosonic system deviations appear.

\begin{figure}
\begin{center}
\includegraphics[angle=0, width=0.8\columnwidth]{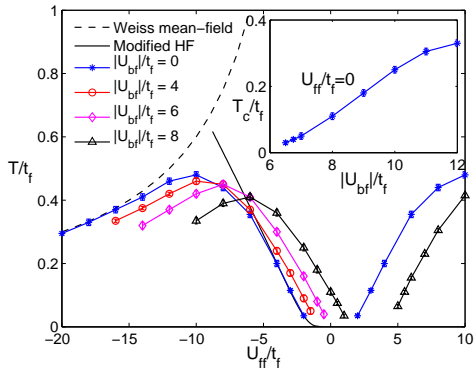}
\caption{(Color online) S-wave superfluid (left) and antiferromagnetic ($U_{\rm bf}=0$ and $\vert U_{\rm bf}\vert / t_{\rm f} = 8$ on the right) phase transition of the Bose-Fermi Hubbard model on the $3d$ cubic lattice at filling $n_{\rm b}=1$ and $n_{\uparrow}=n_{\downarrow}=1/2$ and with $U_{\rm bb}/t_{\rm f}=20$ and $t_{\rm b}/t_{\rm f}=1$ for different boson-fermion interactions $U_{\rm bf}$. 
%The bosonic parameters are $U_{\rm bb}/t_{\rm f}=20$ and $t_{\rm b}/t_{\rm f}=1$.
The DMFT results interpolate between the Weiss mean-field result  and the T-matrix/BCS result ('modified HF') (see text). Inset: Critical temperature for pairing for non-interacting fermions ($U_{\rm ff}=0$) as a function of the boson-fermion interaction $U_{\rm bf}$.  The transition temperature is exponentially suppressed at low $U_{\rm ff}$ for all $U_{\rm bf}$ (not shown). 
%Detecting a possible d-wave superfluid~\cite{Mathey06} is beyond the scope of this work.
}
\label{Fig_phase_diagram_Ubf}
\end{center}
\end{figure}

At half filling the pure fermionic system ($U_{\rm bf}=0$) exhibits particle-hole symmetry:  The superfluid phase transition on the attractive side $U_{\rm ff}<0$ is mirror reflected around $U_{\rm ff}=0$ into a  anti-ferromagnetic transition on the repulsive side, $U_{\rm ff}>0$, as is shown in Fig.~\ref{Fig_phase_diagram_Ubf} (although both have SU(2) character, we already use the terminology appropriate for $U_{\rm bf} \neq 0$). The DMFT results interpolate between the Weiss mean-field result $T_{\text{MF}}=6t^2/|U_{\rm ff}|$ valid at strong coupling and the T-matrix/BCS result at weak coupling~\cite{Keller01}. We first study how the superfluid and anti-ferromagnetic phase transition are affected by the presence of strongly condensed bosons with a speed of sound exceeding the Fermi velocity (referred to as fast bosons), and focus on the  $s$-wave pairing transition. The bosons can then be treated in the Bogoliubov approximation \cite{Viverit} and the effective interaction between the fermions is given by
\begin{eqnarray}
U^{\rm eff}_{\rm ff}(\mathbf{k}, \omega) & = &  U_{\rm ff} + U_{\rm bf}^2 \chi_0(\mathbf{k}, \omega) \label{eq:induced_interaction} \\
 & = & U_{\rm ff} +  \frac{U_{\rm bf}^2 2 n_{\rm b} (zt_{\rm b} + \epsilon^{\rm b}_{\mathbf{k}}) }{ \omega^2 - (zt_{\rm b} + \epsilon^{\rm b}_{\mathbf{k}}) ( (zt_{\rm b} + \epsilon^{\rm b}_{\mathbf{k}}) + 2 n_{\rm b} U_{\rm bb} ) }, \nonumber
\end{eqnarray}
with $\chi_0(\mathbf{k}, \omega)$ the density-density response function. With a strong condensate, the zero temperature expression can be used since the Bose condensation temperature is much higher than the BCS temperature. When the bosonic sound velocity $s_{\rm b} = (2 n_{\rm b} U_{\rm bb} t_{\rm b} )^{1/2}$ is much higher than the Fermi velocity, retardation effects are negligible~\cite{Viverit} and the induced interaction is always attractive on the Fermi sphere. 
The induced interaction is then $U^{\rm ind}(\mathbf{k}) = -\frac{U_{\rm bf}^2}{U_{\rm bb}}  c_1(\mathbf{k}) = -\frac{U_{\rm bf}^2}  {U_{\rm bb}} \frac{1}{1 + \xi^2(z-\sum_{j=1}^d \cos(k_ja) )}$ (with $\xi = \sqrt{t_b / 2n_bU_{\rm bb}}$ the healing length), and an on-site effective interaction $U^{\rm eff}_{\rm ff} = U_{\rm ff} -\frac{U^2_{\rm bf}}{U_{\rm bb}} \sum_{\mathbf{k}} c_1(\mathbf{k})$ is found. The effective hopping follows from a mean-field decoupling of the nearest neighbor interaction and is $t_{\rm f}^{\rm eff} = t_{\rm f} - \frac{U^2_{\rm bf}}{U_{\rm bb}} \langle c^{\dagger}_{i \sigma} c_{j \sigma} \rangle \sum_{\mathbf{k}}  c_1(\mathbf{k}) \cos(k_x)$.

This leads to the phase diagram shown in Fig.~\ref{Fig_phase_diagram_Ubf} where for small $\vert U_{\rm ff} \vert$ s-wave pairing is enhanced by stronger boson-fermion interactions and anti-ferromagnetism is suppressed. Pairing can hence occur for non-interacting and repulsive pure fermions. The inset of Fig.~\ref{Fig_phase_diagram_Ubf} shows that the transition temperature in the purely induced case ($U_{\rm ff} = 0$) can be of the same order as for an attractive fermionic system without bosons~\cite{Viverit}. This holds even for values of $n_{\rm b} U_{\rm bb}$ far outside the Bogoliubov regime. For stronger interspecies interactions than the ones shown, phase separation occurs~\cite{Molmer98, Viverit00, Buechler03} which prevents a further increase of $T_c$. 
%In a cold atom experiment, such a phase will be hard to create for entropic reasons. The bosons with densities $0.05 \le n \le 0.95$ will be condensed and have almost no entropy because the condensation temperature can easily be 10 times higher as the BCS temperature. The entropy of the system will thus almost be entirely carried by the fermions.

\begin{figure}
\begin{center}
\includegraphics[angle=0, width=0.8\columnwidth]{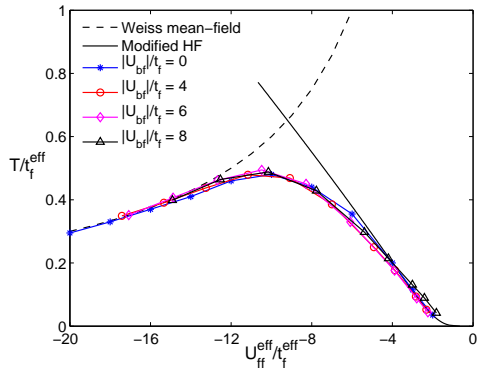}
\caption{(Color online) The phase diagrams of Fig.~\ref{Fig_phase_diagram_Ubf} can be collapsed onto the phase diagram of a pure fermionic model with renormalized hoppings $t_{\rm f}^{\rm eff} = t_{\rm f} - c'_2 U_{\rm bf}^2/U_{\rm bb} $ and on-site repulsions $U_{\rm ff}^{\rm eff} = U_{\rm ff} - c'_1 U_{\rm bf}^2/U_{\rm bb} $, with $c'_1$ and $c'_2$ fitting constants.  Error bars are of the order of the symbol size and omitted for clarity. }
\label{Fig_phase_diagram_collapse}
\end{center}
\end{figure}

The shape of the phase boundary for large $U_{\rm bf}$ in Fig.~\ref{Fig_phase_diagram_Ubf} looks surprisingly similar to the phase diagram of the purely fermionic system. On the basis of the perturbative arguments given above, we look for effective interactions $U_{\rm ff}^{\rm eff}$ and effective hoppings $t_{\rm f}^{\rm eff}$ of the respective forms  $U_{\rm ff}^{\rm eff} = U_{\rm ff} - c'_1 U_{\rm bf}^2/U_{\rm bb} $ and $t_{\rm f}^{\rm eff} = t_{\rm f} - c'_2 U_{\rm bf}^2/U_{\rm bb} $ with $c'_1$ and $c'_2$ fitting constants. In Fig.~\ref{Fig_phase_diagram_collapse} we see that all transition lines can be collapsed onto each other; {\it i.e.,} that in the presence of a fast bosons self-consistent first order contributions suffice to explain the physics, even far outside the perturbative regime.
An analysis of the quasi-particle weight~\cite{Keller01}, $Z_{\rm qp} = (1 - {\rm Im} \Sigma(i\omega_0) / \omega_0 )^{-1}$ (with $\omega_0 = \pi T$ the lowest Matsubara frequency measured from the Fermi level) in systems where symmetry breaking was disabled confirmed this picture (not shown): Zero quasi-particle weight corresponds to the Mott insulator on the repulsive side ($U_{\rm ff} > 0$ for $U_{\rm bf}=0$) and the molecular density wave on the attractive side. Collapse of the curves with different $U_{\rm bf}$ is observed provided the on-site repulsions, hoppings and $Z_{\rm qp}$ factors are rescaled.

\begin{figure}
\begin{center}
\includegraphics[angle=0, width=\columnwidth]{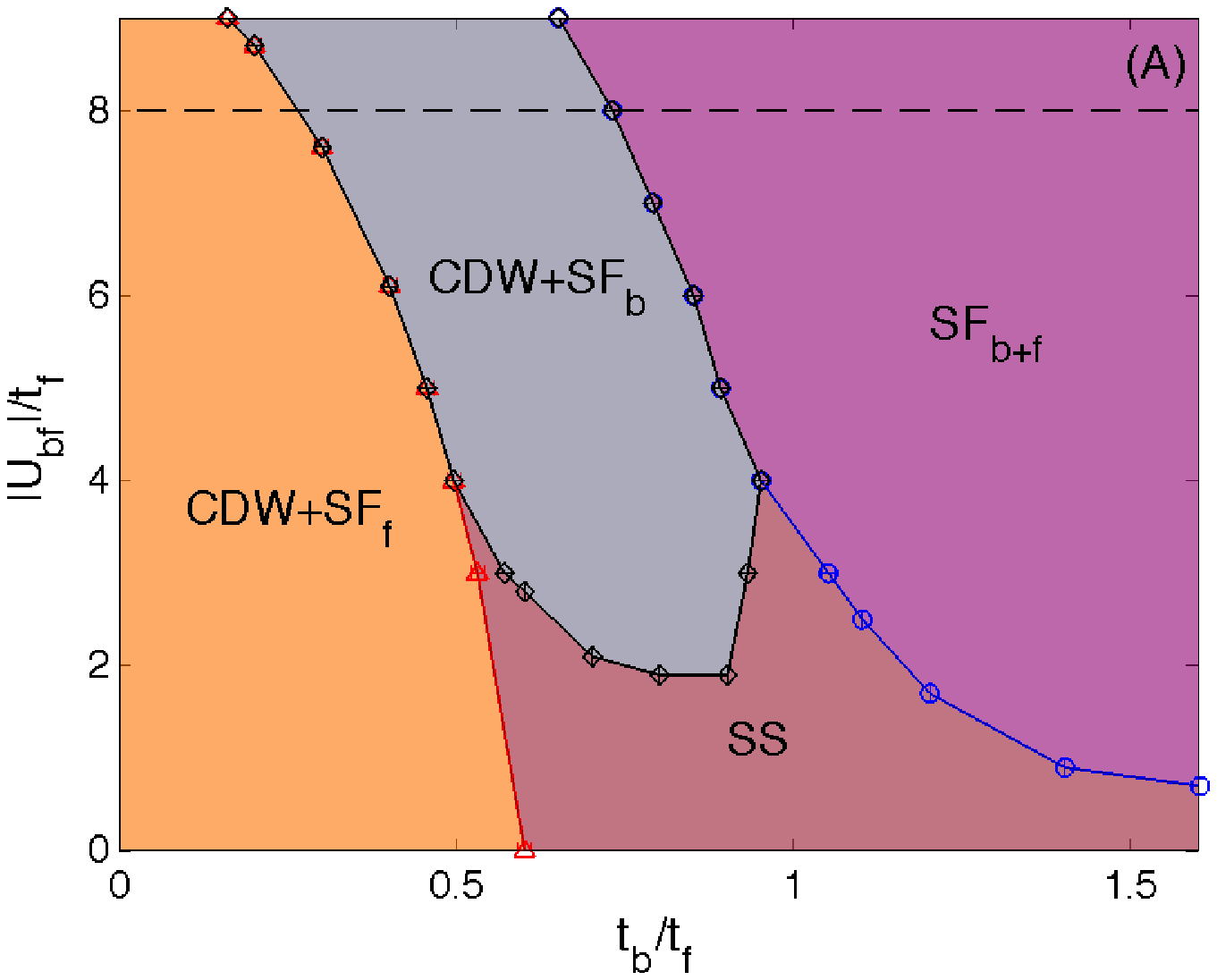}
\includegraphics[angle=0, width=1.0\columnwidth]{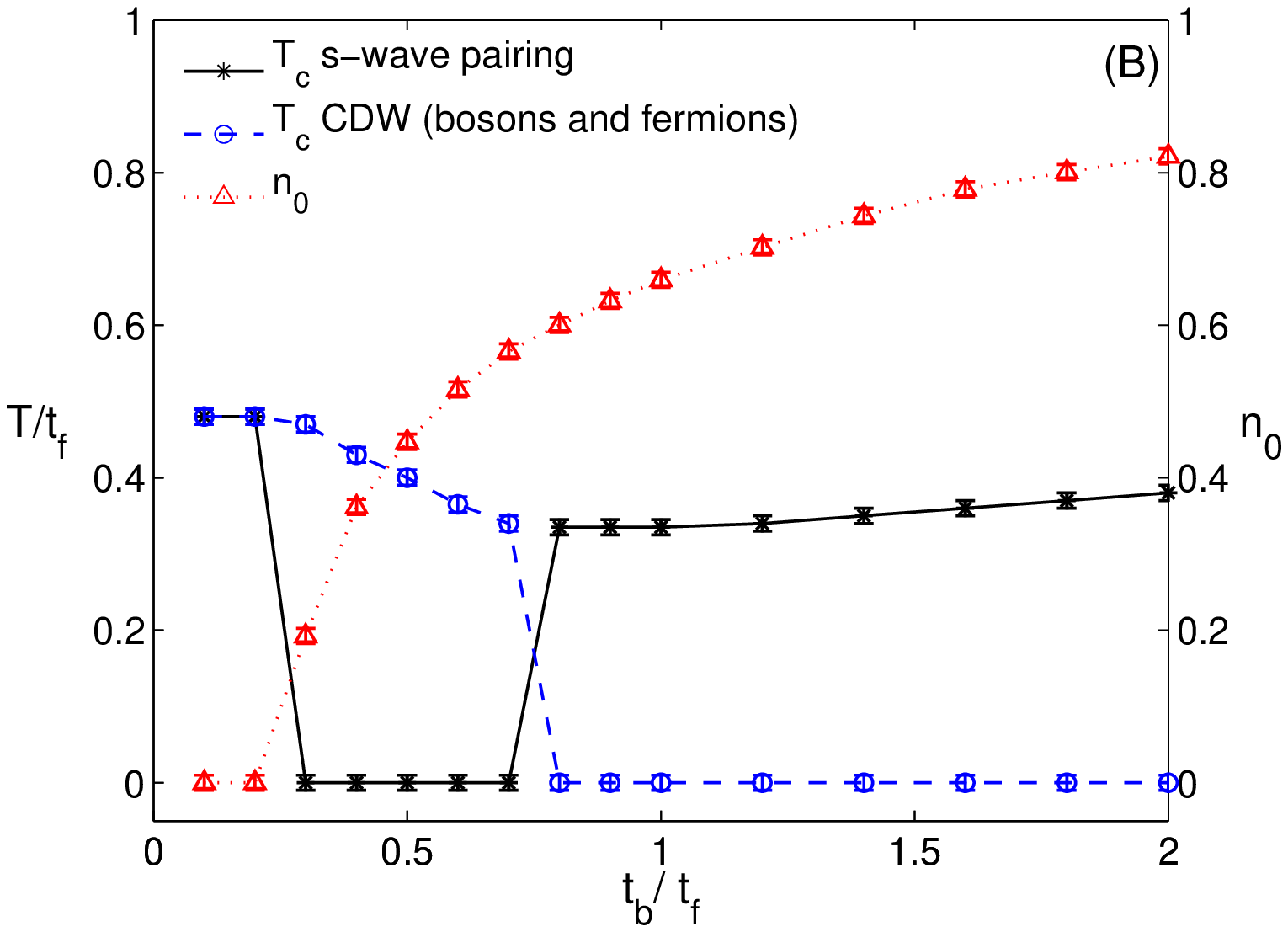}
\caption{(Color online) Panel A: Low temperature phase diagram of the Bose-Fermi Hubbard model on the $3d$ cubic lattice in the $(t_{\rm b}, U_{\rm bf})$ plane. The parameters are $U_{\rm bb}/t_{\rm f}=20, U_{\rm ff}/t_{\rm f}=-10, T/t_{\rm f}=0.2, n_{\uparrow}=n_{\downarrow}=1/2, n_{\rm b}=1$. The labels are: charge density wave (CDW), superfluid fermions (SF$_{\rm f}$), superfluid bosons (SF$_{\rm b}$) and canted supersolid (SS = CDW+SF$_{\rm f}$+SF$_{\rm b}$). The CDW+SF$_{\rm b}$ to SF$_{\rm{b} + \rm{f}}$ and CDW+SF$_{\rm b}$ to CDW+SF$_{\rm f}$ are first order, the other ones are second order. 
Panel B: Critical temperature for s-wave pairing as a function of the bosonic hopping amplitude $t_{\rm b}$ for $|U_{\rm bf}|/t_{\rm f}=8$  (indicated by the dashed line in Fig.~\ref{fig:phase_diagram_Tc_vs_tb}A). 
The bosonic condensate $n_0$ is shown for a temperature corresponding to the maximum of the pairing and charge density wave ordering temperatures.}
\label{fig:phase_diagram_Tc_vs_tb}
\end{center}
\end{figure}

The density-density correlation function in Eq.~(\ref{eq:induced_interaction}) changes dramatically in the absence of a condensate. It may change sign when $\omega$ cannot be set to zero thereby suppressing pairing. This motivates us to  numerically investigate the dependence of the phase transition on the bosonic hopping $t_{\rm b}$,  shown in Fig.~\ref{fig:phase_diagram_Tc_vs_tb}B for strong interactions $U_{\rm ff} / t_{\rm f} = -10$. We see that the system undergoes a sharp first order transition around $t_{\rm b} / t_{\rm f}  \approx 0.75$ between a fermionic superfluid (corresponding to spin singlets in the fermionic spin sector) and a (molecular) charge density wave (corresponding to Neel ordering in the fermionic spin sector). The bosons remain strongly condensed at this point ($n_0 \approx 0.6$), but pick up charge density wave order. The transition temperature varies remarkably little over the different phases, reflecting the underlying $SU(2) \times SU(2)$ symmetry of the pure fermionic model. At very low hoppings ($t_{\rm b} / t_{\rm f} < 0.2$), the bosons become insulating and are very ineffective in influencing the fermions. The fermions can undergo a simultaneous  a pairing and molecular charge order transition, which couples back to the bosons and generates bosonic charge order. We also observed that, except in the close vicinity of a bosonic superfluid-insulator phase transition, bosonic static mean field approximation provides quantitatively correct results in our DMFT scheme, which may be useful for future cluster extensions of this work.
%Hence we neglected $\Delta_b$ in the calculation of the phase diagram shown in Fig.~\ref{fig:phase_diagram_Tc_vs_tb}.
%Although the latter cannot be observed with static mean-field theory, the bosons can always be treated statically far away from the phase transitions.

We repeated this calculation for different values of $U_{\rm bf}$  for a temperature $T/t_{\rm f} = 0.2$ close to the ground state resulting in the phase diagram in the ($U_{\rm bf} , t_{\rm b}$) plane, shown in Fig.~\ref{fig:phase_diagram_Tc_vs_tb}A. 
For large values of $U_{\rm bf}$ we find the same phases as in Fig.~\ref{fig:phase_diagram_Tc_vs_tb}A: a double superfluid, a CDW with a bosonic superfluid, and a CDW with a fermionic superfluid. 

However, for rather low values of $U_{\rm bf}$ and sufficiently large bosonic hoppings we find a supersolid phase, in which bosons and fermions have both types of orderings. In this supersolid, the gaps for pairing and charge order are not equal; this supersolid is a realization of the canted supersolids put forward in Refs~\cite{Mullin71, Liu73}.
The RG study of Ref.~\cite{Mathey06} finds that a $d$-wave superfluid develops for certain parameters in this regime, which may compete with the supersolid. However, seeing such a phase is not possible with single site DMFT. We expect d-wave only to be feasible for low values of $U_{\rm ff}$ and $U_{\rm bf}$ while for large values of $U_{\rm ff}$ and $U_{\rm bf}$ the supersolid is most likely stable.
%It is not unimaginable that this phase is higher in energy than a d-wave paired phase, but our single-site formalism is unable to form d-wave paired phases. It is also in this regime that the renormalization-group study of Ref.~\cite{Mathey06} of the 2d model found a d-wave phase. 
%We also see that Eq.(\ref{eq:induced_interaction}) does not apply to the lowest $U_{\rm bf}$ (when $\vert U_{\rm ff} \vert$ is large), but is for example only valid for $U_{\rm bf} > 4$ in Fig.~\ref{Fig_phase_diagram_Ubf}. 
The transition temperature of the supersolid phase for $|U_{\rm bf}| = 2$ and $t_{\rm b} = t_{\rm f}$ is $T_c \approx 0.48 t_{\rm f}$, rendering an experimental observation with cold gases realistic. This is the same transition temperature as for a supersolid in a bosonic model on a triangular lattice~\cite{Pollet09},  and 50\% higher than the one of an anti-ferromagnet in the 3d Hubbard model~\cite{Fuchs11}. One example of a mixture with promising scattering properties for the supersolid phase  is  $^6$Li-$^7$Li~\cite{Anderson97}.  For low values of $U_{\rm ff}$ the structure of the phase diagram is identical to the one shown in Fig.~\ref{fig:phase_diagram_Tc_vs_tb}A, from which we conclude that the BCS-BEC crossover is not a driving force for the Bose-Fermi Hubbard model at half filling.

%\textcolor{red}{For low values of $U_{\rm ff}$, deep in the BCS side for the pure fermionic model, the phase diagram looks topologically very similar....}

%Next we investigate numerically how the superfluid phase transition depends on the bosonic hopping amplitude $t_{\rm b}$, shown in Fig.~\ref{fig:Tc_vs_tb}. The phase transition for different $t_{\rm b}$ is calculated with the bosons treated in the static mean-field approximation $(\mathbf{\Delta}_{\rm b}(\tau)=0)$ and in the full B-DMFT approximation, with only the latter being properly able to describe the retardation effects in Eq.~(\ref{eq:induced_interaction}) which become important when the bosonic condensation temperature goes down. We see in Fig.~\ref{Fig_phase_diagram_tb} that the retardation effects can be quite large (and non-monotonous with $U_{\rm bf}$, as explained below) for $t_{\rm b} / t_{\rm b} = 0.5$. For immobile bosons, $t_{\rm b} = 0$ (atomic limit), no condensation occurs and the effect of the bosons on the fermions is just a shift in the chemical potential which does not change the phase diagram. Likewise, when the bosons form a strong Mott insulator ($t_{\rm b} / t_{\rm f} = 0.25$ in Fig.~\ref{Fig_phase_diagram_tb}) their main effect is an unimportant shift in chemical potential.

In conclusion, we developed a single-site DMFT formalism for the Bose-Fermi-Hubbard model allowing for $s$-wave pairing and charge density wave ordering. We computed changes to the pure fermionic phase diagram at fermionic half filling induced by the commensurate bosons, focusing on attractive $U_{\rm ff}$. While fast bosons favor $s$-wave pairing, slow bosons favor charge density order. These different type of instabilities compete, leading to some unexpected phases such as a canted supersolid and the CDW+SF$_{\rm b}$ phase shown in the phase diagram of Fig.~\ref{fig:phase_diagram_Tc_vs_tb}.

{\it Acknowledgments}
Calculations have been performed on the Brutus cluster at ETH Zurich. We acknowledge very helpful discussions with the group of I.~Bloch, the group of T. Esslinger, R. Hulet, N. V. Prokof'ev, and C. Salomon. This project was supported by the Swiss National Science Foundation under Grants No. PZ00P2-121892, PP0022-118866, and by a grant from the Army Research Office with funding from the DARPA OLE program.

\end{document}